\documentclass[%
reprint,
superscriptaddress,
showpacs,
nofootinbib,
amsmath,amssymb,
prc,
floatfix, ]%
{revtex4-1}

\usepackage{color}

\usepackage{graphicx}
\usepackage{dcolumn}
\usepackage{bm}
\usepackage[dvipdfmx,bookmarks=true,colorlinks,%
            citecolor=blue,linkcolor=blue,anchorcolor=blue,filecolor=blue,urlcolor=blue,%
           ]{hyperref}

\allowdisplaybreaks

\newcommand{\bra}[1]{\langle {#1} |}
\newcommand{\ket}[1]{| {#1} \rangle}
\newcommand{\inproduct}[2]{\langle #1 | #2 \rangle}

\begin{document}


\title{
Self-consistent collective coordinate for
reaction path and inertial mass
}

\author{Kai Wen}%
 \email{wenkai@nucl.ph.tsukuba.ac.jp}
 \affiliation{Center for Computational Sciences
              and Faculty of Pure and Applied Sciences,
              University of Tsukuba, Tsukuba 305-8577, Japan}

\author{Takashi Nakatsukasa}%
 \affiliation{Center for Computational Sciences
              and Faculty of Pure and Applied Sciences,
              University of Tsukuba, Tsukuba 305-8577, Japan}
 \affiliation{RIKEN Nishina Center, Wako 351-0198, Japan}

\date{\today}

\begin{abstract}
We propose a numerical method to determine the optimal collective reaction
path for the nucleus-nucleus collision,
based on the adiabatic self-consistent collective coordinate (ASCC) method.
We use an iterative method combining
the imaginary-time evolution and the finite amplitude method,
for the solution of the ASCC coupled equations.
It is applied to the simplest case, the $\alpha-\alpha$ scattering.
We determine the collective path, the potential, and the inertial mass.
The results are compared with other methods,
such as the constrained Hartree-Fock method,
the Inglis's cranking formula, and
the adiabatic time-dependent Hartree-Fock (ATDHF) method.
\end{abstract}

\pacs{24.60.-k, 24.10.Lx, 25.60.Pj, 25.70.Lm}

\maketitle


\section{Introduction}

The time-dependent Hartree-Fock (TDHF) method has been extensively
applied to studies of heavy-ion reaction \cite{Neg82,Sim12,Nak12,MRSU14,NMMY16}.
The TDHF provides a successful description for the time evolution of
one-body observables.
Its small amplitude limit corresponds to
the random-phase approximation \cite{RS80,BR86},
which is a current leading theory for the nuclear response calculations.
However, beyond the linear regime,
it is not trivial to extract the quantum mechanical information from
the TDHF trajectories of given initial values.
It is also well-known that the TDHF has some drawbacks due to its
semiclassical nature \cite{Neg82,RS80}.
For instance, the real-time description of sub-barrier fusion and
spontaneous fission processes is practically impossible,
because a single Slater determinant with a single average mean-field
potential is not capable of describing quantum mechanical processes
in rare channels.

The ``requantization'' of TDHF is a possible solution to these problems,
that was proposed from a view point of the path integral \cite{Rei80,Neg82}.
However, the original quantization prescription requires the identification
of periodic TDHF trajectories, which is a very difficult task.
As far as we know, there have been no application of the theory to realistic
nuclear problems \cite{BSW03}.
A family of the periodic TDHF trajectories is associated with
a collective subspace decoupled from the other intrinsic degrees of freedom.
If we identify the collective subspace
spanned by a small number of canonical variables,
the requantization becomes much easier than finding the periodic orbits
\cite{NMMY16}.
In fact, the theory of the adiabatic TDHF (ATDHF) was aiming at
determining such an optimum collective subspace \cite{BGV76,Vil77,BV78,GR78}.
The ATDHF, however, encounters a ``non-uniqueness'' problem,
namely cannot provide a unique solution for the collective subspace.
In order to uniquely fix the solution,
a prescription, so-called validity condition, was proposed \cite{RG78}.
Goeke, Reinhard, and coworkers have developed a numerical
recipe for the reaction path and inertial mass
solving the ATDHF equations of the initial-value problem \cite{GGR83,RG87}.
Their procedure requires us to calculate a large number of trajectories
with different
initial states, then, to obtain the optimal collective path
as an envelope curve of those \cite{GGR83}.

The self-consistent collective coordinate (SCC) method,
originally proposed by Marumori and coworkers \cite{MMSK80},
is solely based on the invariance principle of the TDHF
equation in the collective subspace,
which treats the collective coordinate $q$ and the momentum $p$
on an equal footing.
The SCC method is able to determine the unique collective path.
In addition, the Anderson-Nambu-Goldstone (ANG) modes are properly
decoupled in the SCC method \cite{Mat86,NMMY16}.
Its weak point is that practical solutions to the basic equation
was restricted to a perturbative expansion around the HF state.
To overcome this perturbative nature of the SCC,
a method treating the coordinate $q$ in a non-perturbative way
but expanding with respect to momenta $p$ has been later proposed.
It is named
``adiabatic self-consistent collective coordinate (ASCC) method''
\cite{MNM00}.
The ASCC provides an alternative practical solution to the SCC \cite{MNM00}:
The state is determined at each value of $q$ by solving the equation
expanded up to the second order in $p$.
The ASCC method has been successfully applied to nuclear structure
problems with large-amplitude shape fluctuations/oscillations
for the Hamiltonian of the separable interactions
\cite{HNMM08,HNMM09,HSNMM10,HSYNMM11,HLNNV12,SHYNMM12,MMNYHS16,NMMY16}.
It should be noted that a solution to the non-uniqueness problem of
the ATDHF was given
by higher-order equations with respect to momenta \cite{MP82,KDW91},
which are similar to the ASCC equations.

In this paper,
we apply the ASCC method to nuclear reaction studies,
then, self-consistently determine the optimal reaction path,
the internuclear potential, and the inertial mass.
Since the separable interactions, such as the pairing-plus-quadrupole
interaction, are not applicable to a system with two colliding nuclei,
we need to treat the Hamiltonian of a non-separable type.
For this purpose, we develop a computer code of a novel numerical technique.
We use a combining procedure of the imaginary-time method \cite{DFKW80}
and the finite-amplitude method \cite{NIY07,AN11,AN13}
for the solution of the ASCC equations.
We show that this method nicely works for the three-dimensional (3D)
coordinate-space representation,
taking a reaction of $^8$Be$\leftrightarrow \alpha+\alpha$
as an example.

The paper is organized as follows.
In Sec.~\ref{sec:theo}, we give the formulation of the basic ASCC
equations to determine the one-dimensional (1D) collective path
and the canonical variables $(q,p)$.
In Sec.~\ref{sec:numer}, we show the numerical results
and compare with those of conventional methods.
Summary and concluding remarks are given in Sec.~\ref{sec:summary}.

\section{\label{sec:theo}Theoretical framework}

\subsection{The adiabatic self-consistent collective coordinate (ASCC) method}

The SCC method aims at determining a collective submanifold
embedded in the large dimensional TDHF space of Slater determinants,
which is maximally decoupled from the remaining intrinsic degrees of freedom.
For the 1D collective path, a pair of canonical variables $(q,p)$ along
the collective path are
introduced by labeling the Slater determinants as $|\psi(p,q)\rangle$,
where $q$ and $p$ respectively represent
the coordinate and the conjugate momentum.
Once the states $\ket{\psi(q,p)}$ are determined,
the (classical) collective Hamiltonian is given by
${\cal H}_\textrm{coll}(q,p)\equiv\bra{\psi(q,p)}\hat{H}\ket{\psi(q,p)}$,
where $\hat{H}$ is the total Hamiltonian of the system.
Therefore, the main task is to determine $\ket{\psi(q,p)}$ on a decoupled
collective submanifold.

In the ASCC \cite{MNM00},
the wave function is written in a form
\begin{eqnarray}
|\psi(p,q)\rangle = e^{i p \hat{Q}(q)}|\psi(q)\rangle,
\label{eq1-1}
\end{eqnarray}
using a local generator $\hat{Q}(q)$ which is defined as
\begin{eqnarray}
\hat{Q}(q)|\psi(q)\rangle = -i \left. \partial_{p}|\psi(q)\rangle
 \right|_{p=0}.
\label{eq1-2}
\end{eqnarray}
The collective coordinate operator $\hat{Q}(q)$ is an
infinitesimal generator of ``accelerating'' the system.
The momentum operator
is introduced in the similar way as an infinitesimal generator for
``translating''
the system, $\hat{P}(q)|\psi(q)\rangle = i \partial_{q}|\psi(q)\rangle$.

Since the Thouless theorem guarantees that small variation of
a Slater determinant can be generated by the particle-hole
(ph) excitations \cite{RS80},
the local generators, $\hat{Q}(q)$ and $\hat{P}(q)$,
can be written in terms of ph and hp operators as
\begin{eqnarray}
\hat{P}(q)&=&i \sum_{n,j}P_{nj}(q)a^{\dag}_{n}(q)a_{j}(q)
+ \mathrm{h.c.}, \label{P2}\\
\hat{Q}(q)&=&\sum_{n,j}Q_{nj}(q)a^{\dag}_{n}(q)a_{j}(q)
+ \mathrm{h.c.}.
 \label{Q(q)}
\end{eqnarray}
In this paper,
the indexes $i, j$ and $n,m$ refer to the hole and particle states
with respect to $\ket{\phi(q)}$, respectively,
Hereafter, the creation and annihilation operators are denoted as
$(a_n^\dagger,a_i)$ instead of $(a_n^\dagger(q),a_i(q))$ for simplicity.
These generators are required to follow the weak canonicity condition
\begin{eqnarray}
 \langle\psi(q)|[i\hat{P}(q),\hat{Q}(q)]|\psi(q)\rangle=1. \label{weak}
\end{eqnarray}

In the ASCC, the collective momentum $p$ is assumed to be small.
Keeping the expansion with respect to $p$ up to the second order,
the invariance principle of TDHF equation
leads to a set of ASCC equations~\cite{MNM00,HNMM07,HNMM08,HSNMM10,
Nak12,NMMY16}
to determine the wave function $\ket{\psi(q)}$
and the local generators $(\hat{P}(q), \hat{Q}(q))$
self-consistently along the collective path.
In this paper, we consider only the 1D collective motion,
without taking the pairing correlations into account.
The equations in the zeroth, first, and second order in momentum
read, respectively,
\begin{eqnarray}
&&\delta \bra{\psi(q)} \hat{H}_{\rm mv}(q)\ket{\psi(q)} = 0
 \label{chf} \\
&&\delta\langle \psi(q)|[\hat{H}_{\rm mv}(q),\frac{1}{i}\hat{P}(q) ]
        - \frac{\partial^{2} V}{\partial q^{2}} \hat{Q}(q)
        |\psi(q)\rangle = 0,
 \label{ASCC1} \\
&&\delta\langle \psi(q)|[\hat{H}_{\rm mv}(q),i\hat{Q}(q)]
         - \frac{1}{M(q)}\hat{P}(q)   |\psi(q)\rangle = 0,
\label{ASCC2}
\end{eqnarray}
where
$\hat{H}_{\rm mv}(q)\equiv \hat{H}-(\partial V/\partial q)\hat{Q}(q)$
is the ``moving'' Hamiltonian.
The collective potential $V(q)$ is defined as
\begin{eqnarray}
V(q)= \langle \psi(q)|\hat{H} |\psi(q)\rangle,
\label{pdef}
\end{eqnarray}
and $M(q)$ is
the inertial mass of the collective motion.
Equation (\ref{chf}) is called
``moving mean-field equation'' (``moving Hartree-Fock (HF) equation''),
and Eqs. (\ref{ASCC1}) and (\ref{ASCC2})
are ``moving random-phase approximation (RPA)''.

In fact, to derive the second-order equation (\ref{ASCC2}), an additional
term called ``curvature term'' \cite{MNM00} is neglected.
Although the exact treatment of the curvature term is possible,
it numerically involves iterative tasks
and has only minor effect on the final result \cite{HNMM07}.
Here, we neglect this curvature term, which leads to Eq. (\ref{ASCC2}).
Equation (\ref{chf}) looks similar to a constrained Hartree-Fock (CHF) equation.
However, the constraint operator $\hat{Q}(q)$ changes along
the collective path $\ket{\psi(q)}$, which is self-consistently
determined by the moving RPA equations (\ref{ASCC1}) and (\ref{ASCC2}).

Substituting $\hat{P}$ and $\hat{Q}$ of Eqs. (\ref{P2}) and (\ref{Q(q)})
into Eqs. (\ref{ASCC1}) and (\ref{ASCC2}) leads to
\begin{eqnarray}
\begin{pmatrix}
A(q) & B(q) \\
B^*(q) & A^*(q)
\end{pmatrix}
\begin{pmatrix}
P(q)\\
P^*(q)
\end{pmatrix}
&=& \frac{\partial^2 V}{\partial q^2}
\begin{pmatrix}
Q(q) \\
Q^*(q)
\end{pmatrix}
\label{ASCC3},\\
\begin{pmatrix}
A(q) & B(q) \\
B^*(q) & A^*(q)
\end{pmatrix}
\begin{pmatrix}
Q(q) \\
-Q^*(q)
\end{pmatrix}
&=& \frac{1}{M(q)}
\begin{pmatrix}
P(q) \\
-P^*(q)
\end{pmatrix}
\label{ASCC4}
,
\end{eqnarray}
where the $A$ and $B$ matrix elements are defined as
\begin{eqnarray}
A_{minj}(q)&=&\langle \psi(q) |a^{\dag}_{i}a_{m} [\hat{H}_{\rm mv}(q),a^{\dag}_{n}a_{j}] | \psi(q) \rangle,
\nonumber\\
B_{minj}(q)&=&-\langle \psi(q) |a^{\dag}_{i}a_{m} [\hat{H}_{\rm mv}(q),a^{\dag}_{j}a_{n}] | \psi(q) \rangle.
\label{abmatrx}
\end{eqnarray}
When all of these matrix elements are real,
Eqs. (\ref{ASCC3}) and (\ref{ASCC4}) can be recast into
an eigenvalue equation
\begin{equation}
\left\{A(q)+B(q)\right\}\left\{A(q)-B(q)\right\}Q(q)
=\omega^{2}(q) Q(q),
\label{ab3}
\end{equation}
with
\begin{equation}
\omega^{2}(q)=\frac{1}{M(q)}\frac{\partial^{2} V}{\partial q^{2}},
\label{ab4}
\end{equation}
where $\omega(q)$ is the moving-RPA eigenfrequency.
$\omega(q)$ can be pure imaginary ($\omega^2(q)<0$).
The generator $\hat{P}(q)$ can be obtained from
a matrix equation for $P_{nj}(q)$,
\begin{equation}
P(q) = M(q) \left\{A(q)-B(q)\right\}Q(q) .
\label{P(q)}
\end{equation}
Equation (\ref{ab3}) has many solutions, among which we choose
the collective mode of our interest.
For instance, in numerical calculation for the scattering
$\alpha+\alpha\leftrightarrow ^8$Be in Sec.~\ref{sec:numer},
the lowest quadrupole mode of excitation is selected.

Since the scale of the coordinate is arbitrary,
the ASCC equations (\ref{chf}), (\ref{ASCC1}), and (\ref{ASCC2})
and the weak canonicity condition (\ref{weak})
are invariant with respect to the scale transformation of the
collective coordinate,
$q\rightarrow \alpha q$ ($p\rightarrow p/\alpha$).
The generators and the collective inertial mass are transformed as
$\hat{Q}(q)\rightarrow \alpha \hat{Q}(q)$,
$\hat{P}(q)\rightarrow \hat{P}(q)/\alpha$, and
$M(q) \rightarrow \alpha^{-2} M(q)$, respectively.
Therefore, when we perform the numerical calculation to
determine the collective coordinate $q$,
we need a condition to fix the scale of the coordinate $q$.
A convenient choice could be the condition that
the mass $M(q)$ is unity,
which we adopt in the present paper.
Then, the eigenvalue $\omega^{2}$ of Eq. (\ref{ab3})
gives the second derivative of $V(q)$ with respect to $q$.

In this way, we obtain a series of states $\ket{\psi(q)}$
on the collective path,
the collective potential $V(q)$ of Eq. (\ref{pdef}), and
the collective inertial mass $M(q)$ equal to unity by
tuning the scale of $q$.
Thus,
the collective Hamiltonian is constructed as
\begin{equation}
{\cal H}_\textrm{coll}=\frac{1}{2}\dot{q}^2 + V(q) .
\label{classical_H_q}
\end{equation}
The canonical quantization of this Hamiltonian immediately leads to
\begin{equation}
\hat{H}_\textrm{coll}=-\frac{1}{2}\left(\frac{d}{dq}\right)^2 + V(q) .
\label{quantum_H_q}
\end{equation}


\subsection{Mapping to different variables}
\label{sec:mapping}

In order to obtain a physical picture of the collective dynamics,
it is often convenient to adopt an ``intuitive'' variable,
such as the distance between two nuclei, $R$.
Of course, the optimal collective coordinate $q$,
determined by the ASCC solutions, is different from $R$, in general.
Nevertheless,
as far as the one-to-one correspondence between $q$ and $R$
is guaranteed,
we may use the variable $R=R(q)$ to modify the scale of the coordinate.
Without affecting the collective dynamics,
the collective Hamiltonian,
Eqs. (\ref{classical_H_q}) and (\ref{quantum_H_q}),
is rewritten in terms of $R$.

Let us denote a new variable as $R$,
defined by the expectation value of
the corresponding one-body Hermitian operator $\hat{R}$.
For instance,
the operator of the relative distance between two symmetric nuclei
($2\times A/2$) is given by
\begin{eqnarray}
\hat{R}\equiv\frac{1}{A/2}
\int d\vec{r} \hat{\psi}^\dagger(\vec{r})\hat{\psi}(\vec{r})
\left\{ z\theta(z)- z\theta(-z) \right\}
,
\label{rdef}
\end{eqnarray}
where $\theta(z)$ is the step function.
We also assume the one-to-one mapping between $q$ and $R$,
$R(q)=\bra{\psi(q)}\hat{R}\ket{\psi(q)}$
and its inverse function $q(R)$.
The transformation of the collective potential,
$V(q)\rightarrow V(R)$, is trivial:
$V(R)=V(q(R))$.
In contrast, the inertial mass is transformed as
\begin{eqnarray}
M(R) =M(q)\left(\frac{dq}{dR}\right)^{2}= \left(\frac{dq}{dR}\right)^{2}
= \left(\frac{dR}{dq}\right)^{-2} ,
\label{mass}
\end{eqnarray}
where we use $M(q)=1$.
The inertial mass $M(R)$ is not constant but depends on $R$.
The collective Hamiltonian is rewritten as
\begin{equation}
{\cal H}_\textrm{coll}^R=\frac{1}{2}M(R)\dot{R}^2 + V(R) .
\label{classical_H_R}
\end{equation}
The quantization identical to Eq. (\ref{quantum_H_q}) is
given by the Pauli's prescription \cite{P33}
\begin{equation}
\hat{H}_\textrm{coll}^R=
-\frac{1}{2}
\frac{1}{\sqrt{M(R)}}\frac{d}{dR}
\frac{1}{\sqrt{M(R)}}\frac{d}{dR}
+ V(R) .
\label{quantum_H_R}
\end{equation}

The mass $M(R)$ requires the calculation of the derivative, $dR/dq$ or $dq/dR$,
in Eq. (\ref{mass}).
These quantities can be obtained by use of
the local generator $\hat{P}(q)$,
\begin{eqnarray}
\frac{dR}{dq} &=&
\frac{d}{dq}\langle \psi(q) |\hat{R}| \psi(q) \rangle
 =\langle \psi(q) | [\hat{R},\frac{1}{i}\hat{P}(q) ] | \psi(q) \rangle
 \nonumber \\
 &=&2\sum_{mi} R_{mi}(q)P_{mi}(q)
,\label{mass3}
\end{eqnarray}
where $R_{mi}(q)$ are the ph matrix elements of $\hat{R}$
with respect to the state $\ket{\psi(q)}$
and assumed to be real.
Since this calculation can be performed using the local quantities at $q$,
it has an advantage over the conventional finite difference,
${dR}/{dq}\approx \{R(q+\delta q)-R(q)\}/{\delta q}$,
with two adjacent points, $\ket{\psi(q)}$ and $\ket{\psi(q+\delta q)}$,
on the collective path.
Thus, we use Eq. (\ref{mass3}) for calculation of the derivatives.

In the present ASCC method,
the variable $R$ is merely a parameter to represent the
collective coordinate $q=q(R)$.
It should be emphasized that this is different from assuming the
collective coordinate as $R$.
First of all, the potential is different,
$V(R)=\bra{\psi(q(R))}H\ket{\psi(q(R))}
\neq \bra{\phi(R)}H\ket{\phi(R)}$.
Here, $\ket{\phi(R)}$ is calculated by minimization
of the total energy with a constraint on $R$.
Even if the state $\ket{\phi(R)}$ is close to $\ket{\psi(q(R))}$,
the inertial masses $M^R(R)$ for the motion along the direction $R$
can be very different from $M(R)$.
The ASCC method guarantees a block-diagonal form of
the inertial tensor between the collective coordinate $q$ and
the rest of intrinsic degrees $\{ \vec{\xi} \}$ perpendicular to $q$.
In contrast, the inertial mass tensor $M^R(R)$
for the coordinate $R$ is not block-diagonal in general.
Thus, we need to adopt its diagonal element $M^R(R)$ which is different
from $M(R)$:
\begin{equation}
M^R(R)-M(R) = \sum_{ij}
{\cal M}_{ij}(q(R)) \frac{d\xi^i}{dR} \frac{d\xi^j}{dR} ,
\end{equation}
where ${\cal M}_{ij}(q(R))$ is the inertial mass tensor
for the intrinsic motion.
Last but not least,
the inertial mass $M^R(R)$ is usually
calculated according to the Inglis's cranking formula,
$M^R_{\rm cr}(R)$
(See Sec.~\ref{sec:CHF+cranking}).
The cranking mass $M^R_{\rm cr}(R)$ cannot take into account the effects of
the time-odd mean fields.
In contrast, the ASCC inertial mass $M(R)$, which is determined from
the moving RPA equation (\ref{ASCC2}), reflects the presence of
the time-odd mean fields.
Therefore, even if the collective coordinates $q$ and $R$ are identical,
the calculated inertial masses may be different.
For instance, for the translational (center-of-mass) motion of the nucleus,
the cranking mass fails to reproduce the total mass,
$M^R_{\rm cr}(R)\neq Am$,
when the effective mass $m^*$ is different from the bare nucleon mass $m$.
It is compensated by the time-odd effect in the ASCC inertial mass,
that leads to the exact relation, $M(R)=Am$.

\subsection{\label{sec:algorithm}Numerical algorithm and details}

\subsubsection{Coordinate-space and mixed representation}

In this paper, we adopt the BKN energy density functional \cite{BKN76}
for the Hamiltonian $\hat{H}$
\footnote{
The diagonal approximation of the center-of-mass energy modifies
the nucleon mass,
$m^{-1} \rightarrow m^{-1}(1-A^{-1})$.
In this paper, we do not adopt this correction,
thus, the nucleon mass is the bare mass.
}.
The BKN energy density functional assumes the spin-isospin symmetry without
the spin-orbit interaction, thus, all the single-particle states
at the HF ground state
are real ($\varphi_i^*(\vec{r}) =\varphi_i(\vec{r})$).
The one-body Hamiltonian is given by
\begin{eqnarray}
h[\rho] &=& -\frac{1}{2m}\nabla^{2}
           + \frac{3}{4}t_{0}\rho(\vec{r})
           + \frac{3}{16}t_{3}\rho^{2}(\vec{r})\nonumber \\
           &&+ \int d\vec{r'} v(\vec{r}-\vec{r'})\rho(\vec{r'}), 
\label{BKN0}
\end{eqnarray}
where $v(\vec{r})$ is the sum of the Yukawa and the Coulomb potentials,
\begin{eqnarray}
v(\vec{r})\equiv V_{0}a\frac{e^{-r/a}}{r} + \frac{(e/2)^{2}}{r}.
\end{eqnarray}
We take the same parameter set as in reference \cite{BKN76}.

For the BKN energy density functional,
it is convenient to utilize the coordinate-space representation.
Each single-particle wave function $\varphi_i(\vec{r})$
is represented in the 3D grid points of the square mesh,
$\varphi_{\vec{k}i}\equiv\varphi_i(\vec{r}_{\vec{k}})$ with
$\vec{r}_{\vec{k}}=\vec{k}\times h=(k_x,k_y,k_z)\times h$,
where $h$ is the mesh size.
Although every quantity is defined locally at $q$,
in this subsection,
we omit the collective coordinate $q$ for simplicity,
such as $\varphi_i(\vec{r};q)\rightarrow \varphi_i(\vec{r})$.
The 3D space is a rectangular
box of volume $10\times10\times16$ fm$^{3}$ with
mesh size $h=0.8$ fm.

We adopt the mixed representation for the moving RPA equation:
The particle-state indices $m$, $n,\cdots$ are replaced by the coordinate
$\vec{r}$.
Thus, the generator $\hat{Q}$ of Eq. (\ref{Q(q)}) is represented as
\begin{eqnarray}
\hat{Q} &= & \int d\vec{r} \sum_j Q_j(\vec{r})
 a^\dagger(\vec{r}) a_j + \mbox{h.c.} \nonumber \\
 &\approx& h^3 \sum_{\vec{k}} \sum_j Q_{\vec{k},j}
 a^\dagger_{\vec{k}} a_j + \mbox{h.c.},
\end{eqnarray}
where $Q_{\vec{k},j}=Q_j(\vec{r}_{\vec{k}})$
and $a^\dagger_{\vec{k}}\equiv a^\dagger(\vec{r}_{\vec{k}})$.
Since the coordinate indices $\vec{r}_{\vec{k}}$ contain
not only the particle states but also hole states,
we should remove the hole parts.
Using the projection operator,
$\hat{C}\equiv 1-\sum_i\ket{\varphi_i}\bra{\varphi_i}$,
this is done by replacing $Q_j(\vec{r})=\inproduct{\vec{r}}{Q_j}$ by
\begin{equation}
\int d\vec{r}' C(\vec{r},\vec{r}') Q_j(\vec{r}')
= Q_j(\vec{r})   - \sum_i \varphi_i(\vec{r}) \inproduct{\varphi_i}{Q_j} ,
\label{projection}
\end{equation}
where $C(\vec{r},\vec{r}')\equiv \delta(\vec{r}-\vec{r}')
-\sum_i \varphi_i(\vec{r})\varphi_i^*(\vec{r}')$.
Equivalently, $\ket{Q_j}$ is replaced by $\hat{C}\ket{Q_j}$.
Similar modification is performed for the generator $\hat{P}$ of Eq. (\ref{P2}).

The matrices of Eq. (\ref{abmatrx}) are represented as
$A_{\vec{k}i\vec{k}'j} =
A_{ij}(\vec{r}_{\vec{k}},\vec{r}_{\vec{k}'})$
and the same for the matrix $B$.
The hole contributions are removed in the same manner as
Eq. (\ref{projection}).
For instance, $(A\cdot Q)_{mi}=\sum_{nj} A_{minj}Q_{nj}$ in
the ph representation becomes
\begin{eqnarray}
(A\cdot Q)_i(\vec{r})
&=& \sum_j \iiint d\vec{r}_1 d\vec{r}_2 d\vec{r}_3 \nonumber \\
&& C(\vec{r},\vec{r}_1)
 A_{ij}(\vec{r}_1,\vec{r}_2) C(\vec{r}_2,\vec{r}_3) Q_j(\vec{r}_3) ,
\end{eqnarray}
which can be discretized as
\begin{equation}
(A\cdot Q)_{\vec{k}i} = h^9
\sum_{j} \sum_{\vec{k}_1\vec{k}_2\vec{k}_3} C_{\vec{k} \vec{k}_1}
 A_{\vec{k}_1 i\vec{k}_2 j} C_{\vec{k}_2 \vec{k}_3} Q_{\vec{k}_3 j} ,
\end{equation}
where $C_{\vec{k}_1 \vec{k}_2}\equiv h^{-3} \delta_{\vec{k}_1\vec{k}_2}
- \sum_i \varphi_{\vec{k}_1 i} \varphi_{\vec{k}_2 i}^*$.

Although we remove the hole-hole contributions in this manner,
the RPA matrices are oversize and contain redundant components.
Therefore, the diagonalization of the moving RPA equation
produces spurious solutions that consist of only hole-hole elements
\footnote{
They should not be confused with the zero-modes
associated with the symmetry breaking of the state $\ket{\psi(q)}$,
which are often called ``spurious modes'' as well.
}.
The number of these spurious modes is equal to square of the
number of the hole orbits, $A^2$.
These spurious solutions are decoupled and have
no influence on physical solutions.
Thus, we simply discard the spurious solutions after the diagonalization
of the RPA matrix.

\subsubsection{Finite amplitude method for the moving RPA solution}
\label{sec:FAM}

Solutions of the moving RPA equation (\ref{ab3})
determine the local generators $\hat{Q}(q)$, then
$\hat{P}(q)$ is obtained from Eq. (\ref{P(q)}).
To evaluate the matrix elements of $A\pm B$ in Eq. (\ref{ab3}),
we adopt the finite amplitude
method (FAM) \cite{NIY07,AN11,AN13,Sto11,LNNM13,HKN13,NKTVR13,PKZX14,KHN15},
especially the matrix FAM (m-FAM) prescription \cite{AN13}.
The FAM requires only the calculations of the
single-particle Hamiltonian constructed
with independent bra and ket states~\cite{NIY07},
providing us an efficient tool
to solve the RPA problem.

Let us assume that the state $\ket{\psi(q)}$ is determined from the moving HF
equation (\ref{chf}).
The single-particle states $\ket{\varphi_i(q)}$ and their energies
$\epsilon_i(q)$ of the hole states are defined by
\begin{equation}
\hat{h}_{\rm mv}(q)
\ket{\varphi_i (q)}= \epsilon_i(q)\ket{\varphi_i (q)} ,
\end{equation}
where $h_{\rm mv}(q)=h_{\rm HF}[\rho_0(q)]-\lambda(q)\hat{Q}(q)$
is the single-particle Hamiltonian reduced from
$\hat{H}_{\rm mv}(q)=\hat{H}-\lambda(q) \hat{Q}(q)$
with $\lambda(q)=\partial V/\partial q$.
The self-consistent density $\rho_0(q)$ is given by
$\rho_0(q)=\sum_i \ket{\varphi_i(q)}\bra{\varphi_i(q)}$.
According to Ref.~\cite{AN13}, the matrix elements,
$(A\pm B)_{\vec{r}i,\vec{r}'j}$ can be calculated as follows:
\begin{equation}
\left( A\pm B \right)_{\vec{r}i,\vec{r}'j} =
\left( h_{\rm mv}(\vec{r},\vec{r}') -\epsilon_i \delta(\vec{r}-\vec{r}')
\right) \delta_{ij}
+ \delta h_{[\vec{r}i,\vec{r}'j]} .
\label{ApmB}
\end{equation}
Here, again, the $q$-dependence is omitted for simplicity.
Using a small real parameter $\eta = 10^{-4}$,
the m-FAM provides the elements
$\delta h_{[\vec{r}i,\vec{r}'j]}$ by
\begin{equation}
\delta h_{[\vec{r}i,\vec{r}'j]}
= \eta^{-1}
\bra{\vec{r}}
\left\{ h_{\rm mv}[\rho_{[\vec{r}'j]}]
- h_{\rm mv}[\rho_0] \right\}
\ket{\varphi_i} ,
\label{FAM_formula}
\end{equation}
where $\rho_{[\vec{r}'j]}$ is defined as
\begin{equation}
\rho_{[\vec{r}j]} = \rho_0+
\eta \left(
\ket{\vec{r}} \bra{\varphi_j}
+ \ket{\varphi_j} \bra{\vec{r}}
\right) .
\end{equation}

Note that Eq. (\ref{FAM_formula}) requires only the operation of
the single-particle Hamiltonian on the hole orbits.
In addition, the single-particle Hamiltonian $h_{\rm mv}$ can be
replaced by the HF single-particle Hamiltonian $h_{\rm HF}$
in Eqs. (\ref{ApmB}) and (\ref{FAM_formula}).
It is trivial, for Eq. (\ref{FAM_formula}), to see the term
$\lambda\hat{Q}$ is canceled by the subtraction.
For Eq. (\ref{ApmB}),
this is because the hole components are always removed
from $\vec{r}$ and $\vec{r}'$, and the generator $\hat{Q}$ has
only ph and hp components.

\subsubsection{Imaginary-time method for the moving HF solution}
\label{sec:ITM}

Let us assume that the $\hat{Q}(q)$ is determined from the moving RPA
equation, and now we want to move to the next point on the collective path
($q\rightarrow q+\delta q$).
This can be done by solving the moving HF equation (\ref{chf}),
with the following constraint:
\begin{equation}
 \bra{\psi(q+\delta q)} \hat{Q}(q) \ket{\psi(q+\delta q)}
 = \delta q ,
\label{step_size_constraint}
\end{equation}
which controls the step size of the collective coordinate $\delta q$.
Equation (\ref{step_size_constraint}) can be understood as
\begin{eqnarray}
&&\bra{\psi(q+\delta q)}\hat{Q}(q) \ket{\psi(q+\delta q)} \nonumber \\
&&= \bra{\psi(q)} e^{i\delta q \hat{P}(q)}
 \hat{Q}(q) e^{-i\delta q \hat{P}(q)} \ket{\psi(q)} \nonumber \\
&&\approx \delta q \bra{\psi(q)}
[ i\hat{P}(q), \hat{Q}(q) ]
\ket{\psi(q)}
=\delta q ,
\end{eqnarray}
by the use of the local generator $\hat{P}(q)$ and
the canonicity condition (\ref{weak}).

Hereafter, in this subsection,
the quantities without the explicit $q$-dependence mean those
defined at $q+\delta q$, such as $\ket{\varphi_i}=\ket{\varphi_i(q+\delta q)}$.
The moving HF equation (\ref{chf}) is iteratively solved using
the imaginary-time method \cite{DFKW80} which is
efficient in the coordinate-space representation.
Each single-particle wave function is evolved as
$\ket{\varphi_i^{(n)}} \rightarrow
\ket{\varphi_i^{(n+1)}} =
\ket{\varphi_i^{(n)}} + \ket{\delta\varphi_i^{(n)}}$,
where
\begin{equation}
\ket{\delta \varphi_i^{(n)}}=-\epsilon \left\{ h_{\rm HF}^{(n)}
        -\lambda^{(n)} \hat{Q}(q) \right\} \ket{\varphi_i^{(n)}} ,
\label{ITM}
\end{equation}
with a small real parameter $\epsilon>0$.
$h_{\rm HF}^{(n)}$ is the single-particle Hamiltonian
calculated with the density $\hat{\rho}^{(n)}
=\sum_i \ket{\varphi_i^{(n)}}\bra{\varphi_i^{(n)}}$.
Here we approximate $\hat{Q}(q+\delta q)$ in equation (\ref{chf})
by $\hat{Q}(q)$,
provided that $\delta q$ is small enough.
The Lagrange multiplier $\lambda$ is determined by the constraint
(\ref{step_size_constraint}).
In the first order in $\epsilon$,
$\lambda^{(n)}$ is given by
\begin{eqnarray}
&&\lambda^{(n)} =
\left(\epsilon \textrm{Tr}\left[ \left\{\hat{Q}(q),\hat{Q}(q)\right\}
 \hat{\rho}^{(n)} \right] \right)^{-1} \times \nonumber \\
&&\left(\delta q -  \textrm{Tr}\left[ \hat{Q}(q) \hat{\rho}^{(n)} \right]
 + \epsilon \textrm{Tr}\left[ \left\{ \hat{Q}(q),h_{\rm HF}^{(n)}\right\} \hat{\rho}^{(n)}
\right] \right),
\label{lambda}
\end{eqnarray}
at each iteration.
Here, the traces are calculated as
\begin{eqnarray}
&&\textrm{Tr}\left[ \left\{ \hat{Q}(q),\hat{Q}(q) \right\} \hat{\rho}^{(n)}
\right] = 2\sum_j
\bra{Q_j(q)}\hat{\rho}^{(n)}\ket{Q_j(q)} \nonumber \\
&&\quad\quad  + 2\sum_{ij}
\bra{\varphi_i(q)}\hat{\rho}^{(n)}\ket{\varphi_j(q)}
\inproduct{Q_j(q)}{Q_i(q)} , \\
&&\textrm{Tr}\left[ \left\{ \hat{Q}(q),\hat{O} \right\} \hat{\rho}^{(n)}
\right] =
\sum_j \bra{Q_j(q)} \hat{O}\hat{\rho}^{(n)} \ket{\varphi_j(q)}
\nonumber \\
&&\quad\quad  +\sum_j \bra{\varphi_j(q)}\hat{O} \hat{\rho}^{(n)} \ket{Q_j(q)}
 + \textrm{c.c.} ,
\end{eqnarray}
with $\hat{O}=1$ and $h_{\rm HF}^{(n)}$.
Note $\ket{Q_j(q)}=\hat{C}(q)\ket{Q_j(q)}$ with
$\hat{C}(q)=1-\sum_i \ket{\varphi_i(q)}\bra{\varphi_i(q)}$.
In actual calculations, we also have constraints on the center of mass
and the direction of the principal axis.
These additional constraints
are easily taken into account by
extending Eqs. (\ref{ITM}) and (\ref{lambda}).

According to Eq. (\ref{chf}), in principle, we should use
$\hat{Q}(q+\delta q)$ in Eq. (\ref{ITM}) instead of $\hat{Q}(q)$,
namely the generator at the same point $q+\delta q$.
The prescription given in Sec.~\ref{sec:ITM} actually approximates the generator
$\hat{Q}(q+\delta q)$ by the one at the previous point $\hat{Q}(q)$.
The approximation significantly reduces the computational task.
This approximation turns out to be very good
as far as $\delta q$ is small enough. After
moving the state from $\ket{\psi(q)}$ to $\ket{\psi(q+\delta q)}$
with small $\delta q$,
the self-consistency can be checked in the following way. 
At $\ket{\psi(q+\delta q)}$ we calculate
the generator $\hat{Q}(q+\delta q)$ by solving the moving RPA equations.
Replacing $\hat{Q}(q)$ by $\hat{Q}(q+\delta q)$ in Eq. (\ref{ITM}) and
changing the constraint condition Eq. (\ref{step_size_constraint}) to
$\bra{\psi(q+\delta q)} \hat{Q}(q+\delta q) \ket{\psi(q+\delta q)} = 0$,
the self-consistency between $\ket{\psi(q+\delta q)}$ and
$\hat{Q}(q+\delta q)$ is guaranteed if
the further imaginary-time evolution of Eq. (\ref{ITM})
keeps the state $\ket{\psi(q+\delta q)}$ invariant.
This is confirmed for the present case.
The validity is also confirmed by the fact that
the final result is invariant
with respect to change of the step size $\delta q$.

\subsubsection{Summary of the numerical algorithm}

We choose the HF ground state as the starting point of the collective path,
$\ket{\psi(q=0)}$.
The HF ground state is always the solution of Eq. (\ref{chf}) with $\partial V/\partial q=0$,
at which the moving RPA equation becomes identical to
the conventional RPA equation.
Therefore, without knowing the generator $\hat{Q}(q)$,
the starting point $\ket{\psi(q=0)}$ can be determined.
The procedure to construct the collective path is given as follows:
\begin{enumerate}
\item Calculate the HF ground state, $\ket{\psi(q=0)}$.
\vspace{-5pt}
\item Solve the RPA equation to obtain
 $\hat{Q}(q=0)$ and $\hat{P}(q=0)$.
\vspace{-5pt}
\item \label{num:moving_HF}
When $\ket{\psi(q)}$, $\hat{Q}(q)$, and $\hat{P}(q)$ are provided,
solve the moving HF equation to obtain the state $\ket{\psi(q+\delta q)}$,
according to the method described in Sec.~\ref{sec:ITM}.
\vspace{-5pt}
\item \label{num:moving_RPA}
Solve the moving RPA equation to obtain the generators,
 $\hat{Q}(q+\delta q)$ and $\hat{P}(q+\delta q)$,
according to the method described in Sec.~\ref{sec:FAM}.
\vspace{-5pt}
\item
Repeat the steps \ref{num:moving_HF} and \ref{num:moving_RPA} to determine
the collective path.
\end{enumerate}
For the step \ref{num:moving_RPA} above, we choose the inertial mass
$M(q)=1$.
Then, the weak canonicity condition (\ref{weak}) determines the scale
of $q$ as
\begin{eqnarray}
2\sum_{mnij} Q_{mi}(q)\left[A(q)-B(q)\right]_{minj}Q_{nj}(q) = 1.
\end{eqnarray}
The scale transformation from $q$ to $R$ is performed by
changing the inertial mass according to Eqs. (\ref{mass})
and (\ref{mass3}).

\subsubsection{Algorithm for fully consistent solutions}

Since $^8$Be is one of the simplest cases,
we also try another method to get the fully self-consistent solutions of
the $\hat{P}(q)$, $\hat{Q}(q)$ and $\ket{\psi(q)}$
that simultaneously satisfy Eqs. (\ref{chf}), (\ref{ASCC1}), and (\ref{ASCC2}).
For $^8$Be, the conventional
constrained calculation on $Q_{20}$ may produce approximate solutions,
$\ket{\psi^{(0)}(q)}$.
Thus, we adopt $\ket{\psi^{(0)}(q)}$ as the initial trial wave functions,
and start the following iteration procedure.
\renewcommand{\labelenumi}{(\roman{enumi})}
\begin{enumerate}
\item Solve the Eqs. (\ref{ASCC1}) and (\ref{ASCC2}) to obtain
    $(\hat{Q}(q),\hat{P}(q))$, by selecting the quadrupole mode $Q_{20}$.
\vspace{-5pt}
\item Use this $\hat{Q}(q)$ to solve the moving HF equation (\ref{chf})
     with the constraint $\bra{\psi(q)} \hat{Q}(q) \ket{\psi(q)}=0$.
\vspace{-5pt}
\item Put the obtained state $\ket{\psi(q)}$ into
     Eqs. (\ref{ASCC1}) and (\ref{ASCC2}), then go back to step (i).
\vspace{-5pt}
\end{enumerate}

We also use the initial trial states prepared by the CHF
calculation with constraint on the relative distance $\hat{R}$.
Although the initial states $\ket{\psi^{(0)}(q)}$ are
different from those obtained with the $\hat{Q}_{20}$ operator,
after the iteration of (i)$-$(iii) converges,
they reach the same self-consistent solutions,
$\ket{\psi(q)}$ and $(\hat{Q}(q),\hat{P}(q))$.

It should be noted again that the prescription in Sec.~\ref{sec:ITM} is
significantly easier than the present iteration (i)$-$(iii).
At every point $q$, the self-consistency requires us to solve
the moving RPA equations many times to determine the self-consistent
state $\ket{\psi(q)}$.
We have confirmed that the solution of these iteration
procedures (i)$-$(iii) is practically identical to the one obtained
with the algorithm in Sec.~\ref{sec:ITM}.

\section{\label{sec:numer}Numerical results}

In this work the BKN energy density functional is adopted as a test for
the numerical application of the ASCC method.
The BKN energy density functional is rather schematic, thus,
we should take the following results in a qualitative sense.

\subsection{\label{sec:RPA}
Results of the RPA calculation at the ground state}

If the frequency $\omega$ is positive ($\omega^2 > 0$) for Eq. (\ref{ab3}),
we may construct the normal-mode excitation operator $\Omega^\dagger(q)$
from the generators $(\hat{Q}(q),\hat{P}(q))$ as
\begin{equation}
\Omega^\dagger(q)=\sqrt{\frac{\omega(q)}{2}} \hat{Q}(q)
 -\frac{i}{\sqrt{2\omega(q)}} \hat{P}(q) .
\end{equation}
For a Hermitian one-body operator $\hat{D}$, defined by Eq. (\ref{Q(q)})
with replacement of $Q_{nj}(q)\rightarrow D_{nj}$,
the transition matrix element is given by
\begin{eqnarray}
\bra{\omega}\hat{D}\ket{0} &\equiv&
 \bra{0}[\Omega(q), \hat{D}]\ket{0}
 \nonumber \\
 &=& \sqrt{\frac{2}{\omega}} \sum_i\int d\vec{r}
     P_i (\vec{r}) D_i(\vec{r}) .
\label{trmx}
\end{eqnarray}
We assume that, for the coordinate representation
of the operator such as $P_i(\vec{r})$ or $D(\vec{r})$,
the hole components are always
projected out according to Eq. (\ref{projection}).
The collective character of the state $\ket{\omega}$ can be identified
by choosing the one-body operator $\hat{D}$.
For instance, the translational motion along $z$ axis is identified by
a sizable transition matrix element of the center-of-mass operator,
$\hat{D}=A^{-1}\sum_k^A z_k$.
For the relative motion of two-alpha particles, we may choose
the mass quadrupole operator
(Sec.~\ref{sec:RPA_8Be}).


\subsubsection{\label{sec:alpha}
Translational motion of a single $\alpha$ particle}

First, we show results for the single $\alpha$ particle.
In this case, the model space is a sphere of radius $R = 7$ fm
with various mesh sizes $h=0.5\sim 1.4$ fm.
Note that the ground state of the system
is a trivial solution of the ASCC equation (\ref{chf}).
We can clearly identify the three translational modes for
$x$, $y$, and $z$ directions, degenerated in energy
at $\omega_{\rm com} \leq 1$ MeV.
Using smaller mesh size,
the eigenfrequency of the translational motion approaches to zero.
There are no low-lying excited states in the $\alpha$ particle
because of its compact and doubly-closed characters.
The calculated energy of the lowest excited state
is larger than 20 MeV.

\begin{figure}
\begin{centering}
\includegraphics[width=0.90\columnwidth]{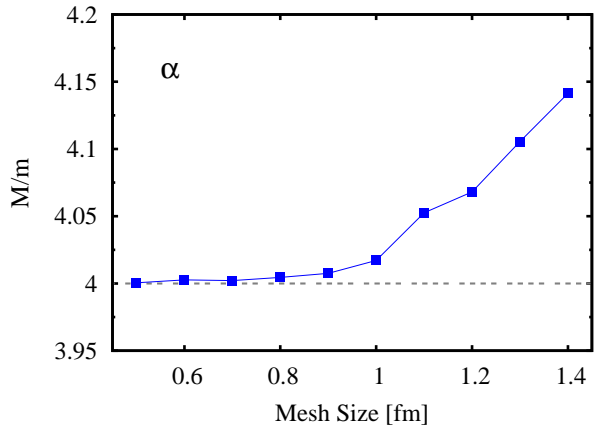}
\par\end{centering}
\caption{\label{fig:massa}(Color online)
Calculated translational mass of the $\alpha$ particle
in units of nucleon's mass $m$,
}
\end{figure}

Using Eqs. (\ref{mass}) and (\ref{mass3}) with $R$ as
the center of mass, we
calculate the inertial mass of the translational motion
of the $\alpha$ particle.
Figure \ref{fig:massa} shows the results calculated with
different mesh size $h$ of the 3D grid.
Since this is the trivial center-of-mass motion
of the total system,
this should equal the total mass, $M=Am$ with $A=4$.
As the mesh size decreases,
the total mass certainly converges to the value of $4m$.
In the following, we adopt
the mesh size $h=0.8$ fm.

\subsubsection{\label{sec:RPA_8Be}
Relative motion of two $\alpha$ particles in $^{8}$Be}

\begin{figure}
\includegraphics[width=0.90\columnwidth]{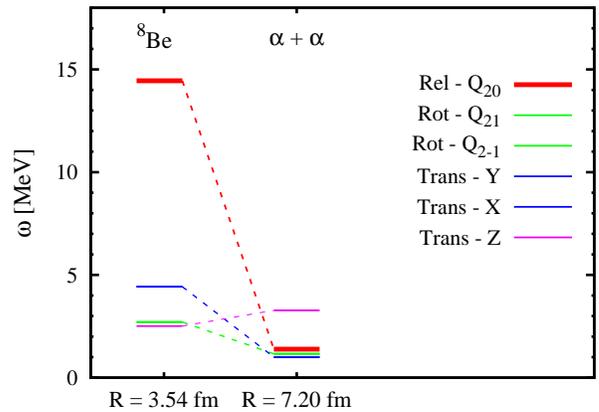}
\caption{\label{fig:eigenfre}(Color online)
Calculated eigenfrequencies for the ground state of $^{8}$Be (left column)
and the two well-separated $\alpha$'s at distance $R = 7.2$ fm (right column).
The three modes of translational motion and two modes of rotational motion
are shown by thin lines, while the thick line indicates
the $K=0$ quadrupole oscillation. 
The translational motions along the $x$ and the $y$ directions
are degenerate in energy, and the same for the rotational motions.
}
\end{figure}

\begin{figure}
\includegraphics[width=0.90\columnwidth]{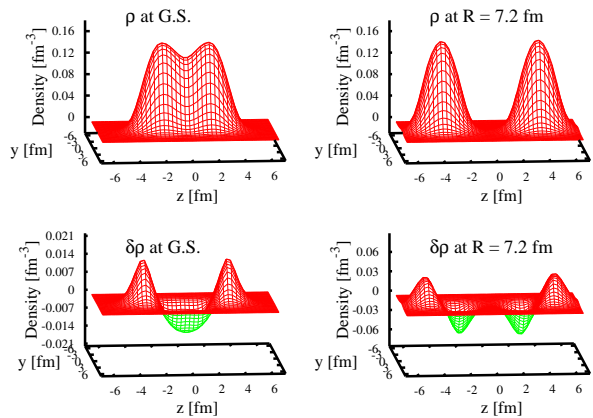}
\caption{\label{fig:tdens}(Color online)
The density distribution $\rho(\vec{r})$ for $^8$Be
in the upper panels,
and the transition density $\delta\rho(\vec{r})$ of the lowest
mode of excitation in the lower panels.
The left panels show those at the ground state and the right at
$R = 7.2$ fm.
Those on the $y-z$ plane are plotted.
}
\end{figure}

Figure \ref{fig:eigenfre} shows the calculated eigenfrequencies
for the ground state of $^{8}$Be
and the two well separated $\alpha$'s at distance $R = 7.2$ fm.
Since the ground state of $^8$Be is deformed,
there appear the rotational modes of excitation as the zero modes,
in addition to the three independent modes of the translational motion.
Because of the axial symmetry of the ground state,
the rotation about the symmetry axis ($z$ axis) does not appear.
In Fig. \ref{fig:eigenfre} the calculation produces
two rotational modes of excitation around 2.8 MeV
with large transition matrix element of the $K = 1$ quadrupole
operator,
$\hat{Q}_{2\pm 1}\equiv
\int r^2 Y_{2\pm 1}(\hat{r})\hat{\psi}^\dagger(\vec{r})\hat{\psi}(\vec{r}) d\vec{r}$.
The finite energy of these rotational modes comes from the finite
mesh size discretizing the space.
Besides these five zero modes,
the lowest mode of excitation turns out to have a sizable
transition strength of the $K=0$ quadrupole operator
$\hat{Q}_{20}\equiv
\int r^2 Y_{20}(\hat{r})\hat{\psi}^\dagger(\vec{r})\hat{\psi}(\vec{r})
 d\vec{r}$.
This mode corresponds to the elongation of $^8$Be.
The transition density is given by
\begin{eqnarray}
\delta \rho(\vec{r}) &\equiv& \bra{\omega}
 \hat{\psi}(\vec{r})\hat{\psi}^\dagger(\vec{r})\ket{0}
= \bra{0} \left[ \Omega,
 \hat{\psi}(\vec{r})\hat{\psi}^\dagger(\vec{r})\right]\ket{0}
\nonumber \\
&=& \sqrt{\frac{2}{\omega}}\sum_i P_i(\vec{r}) \varphi_i(\vec{r}).
\end{eqnarray}
The left panels of Fig.~\ref{fig:tdens} show the density profile of $^8$Be
and the transition density $\delta \rho(r)$
corresponding to the lowest RPA normal mode.
We can see an elongated structure along the $z$ direction
in the ground state.
The lowest mode of excitation corresponds to the
change of its elongation ($\beta$-vibration).

We also perform the same calculation for the state in which
two $\alpha$ particles are located far away,
at the relative distance $R=7.2$ fm.
In the right panel of Fig. \ref{fig:tdens},
we clearly see that the two $\alpha$ particles are well separated,
and the quadrupole mode in fact corresponds to the translational motion
of the $\alpha$ particles in the opposite directions,
namely, the relative motion of two $\alpha$'s.
The excitation energy almost vanishes for this normal mode
(Fig. \ref{fig:eigenfre}).

\subsection{\label{sec:ASCC_result} Results of the ASCC method}

In Sec.~\ref{sec:RPA_8Be},
we show that the the lowest quadrupole mode of excitation at the
ground state of $^{8}$Be may change its character and lead to the relative motion
of two $\alpha$'s at the asymptotic region.
We adopt this mode as the generators $(\hat{Q}(q),\hat{P}(q))$
of the collective variables $(q,p)$, then,
construct the collective path.

\subsubsection{Collective path, potential, and inertial mass}

\begin{figure}
\begin{centering}
\includegraphics[width=0.90\columnwidth]{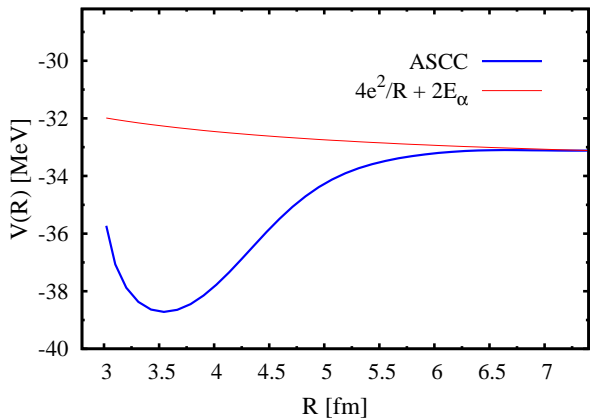}
\par\end{centering}
\caption{\label{fig:pot3}(Color online)
Potential energy as a function of the relative
distance $R$.
The solid (blue) line corresponds to $V(R)$ on the ASCC collective
path, while the dashed (red) line shows
$4e^{2}/R+2E_\alpha$ for reference.
}
\end{figure}

We successfully derive the collective path $\{ \ket{\psi(q)};
 q=0,\delta q, 2\delta q, \cdots \}$
connecting the ground state of $^8$Be into the well-separated two $\alpha$
particles.
The inertial mass $M(q)$ is taken as unity and
the collective potential is calculated according to Eq. (\ref{pdef}).
Then, according to Sec.~\ref{sec:mapping},
the collective coordinate $q$ is mapped onto the relative distance
$R\equiv\bra{\psi(q)}\hat{R}\ket{\psi(q)}$ with Eq.~(\ref{rdef}).
Figure \ref{fig:pot3} shows the obtained potential energy
along the ASCC collective path.
As a reference,
we also show the pure Coulomb potential between two $\alpha$ particles
at distance $R$, $4e^{2}/R+2E_\alpha$, where $E_\alpha$ is the
calculated ground state energy of the isolated $\alpha$ particle.
Apparently, it asymptotically approaches the pure Coulomb potential.
As two $\alpha$'s get closer,
the potential starts to deviate from the Coulomb potential at $R<6$ fm
and finally reaches the ground state of $^8$Be.
The ground state is at $R = 3.54$ fm, and
the top of the Coulomb barrier is at $R = 6.6$ fm.
Note that the path is determined self-consistently without any
{\it a priori} assumption.

\begin{figure}
\begin{centering}
\includegraphics[width=0.90\columnwidth]{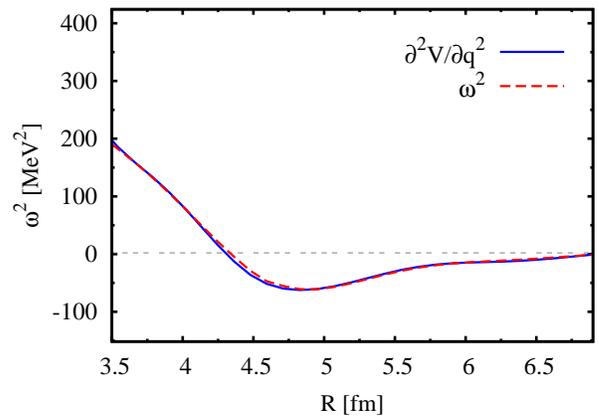}
\par\end{centering}
\caption{\label{fig:checkas}(Color online)
$\omega^{2}$ in Eq. (\ref{ab3}) and
$\partial^{2}V/\partial q^{2}$
of the ASCC calculation
as a function of relative distance $R$.
}
\end{figure}

With this calculated potential, we may
check the self-consistency of the ASCC potential and
the eigenfrequency.
If the collective path perfectly follows
the direction defined by the local generators $(\hat{Q}(p),\hat{P}(q))$
at each point of $q$,
the second derivative of the potential $d^2V/dq^2$
should coincide with the eigenfrequency $\omega^2$ of the moving RPA equation.
The almost perfect agreement between these is shown in Fig. \ref{fig:checkas}.

For the region of $R<3.5$ fm,
there exists some discrepancy between $d^2V/dq^2$
and $\omega^2$.
In this region, the $^8$Be nucleus has even more compact
shapes than the ground state, then,
the coordinate $q$ and $R$ become almost orthogonal to each other,
losing the one-to-one correspondence between them.
In other words, the states $\ket{\psi(q)}$ change as $q$ gets smaller,
but keep $R=\bra{\psi(q)}\hat{R}\ket{\psi(q)}$ almost constant.
In addition, the moving RPA frequency $\omega$ becomes larger than
the particle threshold energy, entering in the continuum.
Thus, in this region of $R<3.5$ fm,
the results somewhat depend on the adopted box size.

\begin{figure}
\begin{centering}
\includegraphics[width=0.90\columnwidth]{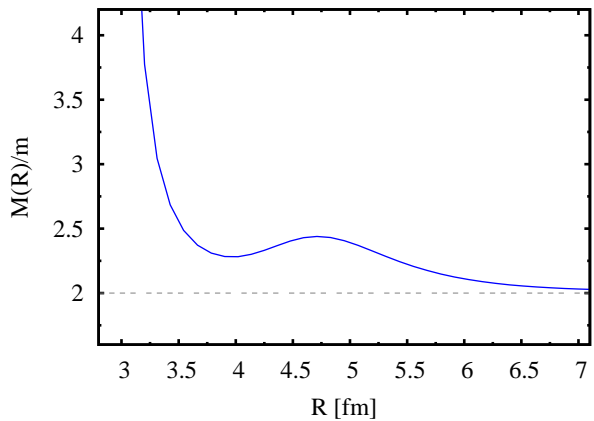}
\par\end{centering}
\caption{\label{fig:mass14}(Color online)
Inertial mass in units of the nucleon's mass $m$
for the collective path of $\alpha+\alpha \leftrightarrow ^{8}$Be,
as a function of the relative distance $R$.
}
\end{figure}

Figure \ref{fig:mass14} shows the obtained
inertial mass $M(R)$ as a function of $R$ for the scattering between two $\alpha$'s 
As the two $\alpha$'s are far away,
the ASCC inertial mass asymptotically produces the exact reduced mass of $2m$.
This means that the collective coordinate $q$ becomes parallel to the relative
distance $R$, even though we do not assume so.
At $R < 3.54$ fm, the value of inertial mass $M(R)$ increases.
This is due to the decrease of the factor $dR/dq$ in Eq. (\ref{mass}).
Making the system even more compact than the ground state,
$M(R)$ rises up drastically,
which means that the coordinates $q$ and $R$ become almost orthogonal.

\subsubsection{Phase shift for $\alpha-\alpha$ scattering}

\begin{figure}
\begin{centering}
\includegraphics[width=0.90\columnwidth]{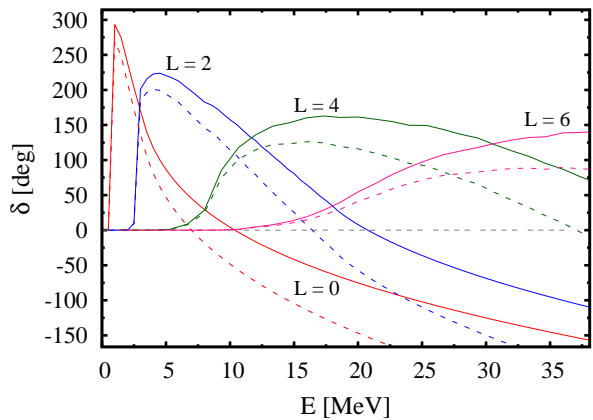}
\par\end{centering}
\caption{\label{fig:phsh}(Color online)
Nuclear phase shift for the scattering between two $\alpha$ particles,
 as a function of incident
energy $E$.
The solid lines indicate the results obtained with the ASCC inertial mass
$M(R)$, while the dashed lines
are calculated with the constant reduced mass 2$m$.
}
\end{figure}

The ASCC calculation provides us the collective Hamiltonian
along the optimal reaction path.
Using this, we demonstrate the calculation of nuclear phase shift.
We should take this result in a qualitative sense,
because of a schematic nature of the BKN energy density functional.

Using the collective potential $V(R)$ and the inertial mass $M(R)$
obtained in the ASCC calculation,
the nuclear phase shift for the angular momentum $L$
at incident energy $E$ is calculated in the WKB approximation
as \cite{Bri85,SKF83}
\begin{eqnarray}
\delta_L(E)=\int_{R_{0}}^{\infty}k(R) dR - \int_{R_c}^{\infty}k_{c}(R) dR,
\end{eqnarray}
with
\begin{eqnarray}
k^2(R) &=& 2M(R)\left\{E-V(R)-\frac{\left(L+\frac{1}{2}\right)^{2}}{4mR^{2}}\right\},\nonumber\\
k_c^2(R) &=& 4m\left\{E-\frac{4e^{2}}{R}-\frac{\left(L+\frac{1}{2}\right)^{2}}{4mR^{2}}\right\} ,
\end{eqnarray}
where $k(R)$ and $k_c(R)$ are the wave numbers in the radial motion with and
without the nuclear potential.
$R_{0}$ and $R_{\rm c}$ are the outer turning points for
the potentials $V(R)$ and ${4e^2}/{R}$, respectively,
i.e. $k(R_0)=k_c(R_c)=0$.
The centrifugal potential is approximated as
$(L+1/2)^2/(2\mu R^{2})$
with the reduced mass $\mu=2m$ and
the semiclassical approximation for $L(L+1)$.
We assume $V(R)=+\infty$ for $R<3$ fm
in which the obtained optimal reaction path is almost orthogonal to $R$.

Figure \ref{fig:phsh} shows the calculated nuclear phase shifts for the
scattering between two $\alpha$'s.
The dashed line is calculated with the same potential $V(R)$
but with the constant reduced mass, $M(R)\rightarrow \mu=2m$.
We can see
the prominent increase of the nuclear phase shift caused
by the coordinate-dependent ASCC inertial mass $M(R)$.
We should remark that the energy of the resonance in $^8$Be is not
reproduced with the BKN energy density functional.
In fact, the present calculation leads to the stable ground state for $^8$Be;
$E({^8{\rm Be}}) < 2 E_\alpha$.
Thus, we should regard this result as a qualitative one.
Nevertheless, the
basic features of phase shifts for the $\alpha-\alpha$ scattering
are roughly reproduced.
This demonstrates the usefulness of the requantization using the ASCC
calculation.

\subsection{Comparison with other approaches}

We compare the present ASCC results with those obtained with other
approaches:
(i) CHF + cranking inertia,
(ii) CHF + local RPA,
and (iii) ATDHF.
We adopt the same model space as the ASCC calculations for these calculations.
For the constraint operators of CHF calculation in (i) and (ii),
we adopt the $K=0$ mass quadrupole operator
$\hat{Q}_{20}$ and the relative distance $\hat{R}$.

\subsubsection{CHF + cranking inertia}
\label{sec:CHF+cranking}

Since $^8$Be is the simplest system and has a prominent
$\alpha+\alpha$ structure even at the ground state,
the collective path can be approximated by
more conventional CHF calculations
with a constraint operator as either $\hat{Q}_{20}$ or $\hat{R}$.
The potential is defined as $V_{\rm CHF}(R)
=\bra{\psi_{\rm CHF}(R)} \hat{H} \ket{\psi_{\rm CHF}(R)}$.
For the inertial mass,
the Inglis's cranking formula is widely used.
There are two kinds of cranking formulae:
The original formula is derived by the adiabatic perturbation, which
is given for the 1D collective motion as
\begin{equation}
M_{\rm cr}^{\rm NP}(R)=2 \sum_{m,i}
\frac{|\bra{\varphi_m(R)}\partial/\partial R\ket{\varphi_i(R)}|^2}
{e_m(R)-e_i(R)} ,
\label{NP_cranking}
\end{equation}
where the single-particle states and energies are defined with
respect to $h_{\rm CHF}(R)=h_{\rm HF}[\rho]-\lambda(R) \hat{O}$ as
\begin{equation}
h_{\rm CHF}(R)\ket{\varphi_\mu(R)}=e_\mu(R))\ket{\varphi_\mu(R)} ,
\quad \mu = i, m .
\end{equation}
Note that, depending on choice of the constraint operator,
$\hat{O}=(\hat{Q}_{20},\hat{R})$,
we obtain slightly different $\ket{\varphi_i(R)}$ even at the same $R$.

Another formula, which is more frequently used in many applications
and also called the cranking inertial mass,
is derived, by assuming the separable interaction and taking
the adiabatic limit of the RPA inertial mass,
\begin{equation}
M_{\rm cr}^{\rm P}(R)=
\frac{1}{2} \left\{ S^{(1)}(R)\right\}^{-1}S^{(3)}(R)\left\{S^{(1)}(R)\right\}^{-1} ,
\label{P_cranking}
\end{equation}
with
\begin{equation}
S^{(k)}(R)=\sum_{m,i}\frac{|\bra{\varphi_m(R)}\hat{R}\ket{\varphi_i(R)}|^{2}}
{\{e_{m}(R)-e_{i}(R)\}^{k}}.
\label{S_k}
\end{equation}
The residual fields induced by the density
fluctuation is neglected in both of these cranking formulae.
According to Ref.~\cite{Bar11}, we call the former one in Eq. (\ref{NP_cranking})
``non-perturbative'' cranking inertia and the latter in Eq. (\ref{P_cranking})
``perturbative'' one.
The method of CHF + cranking inertia has been widely used for
many applications, including
studies of nuclear structure
\cite{BK68-1,BK68-2,YLQG99,PR09,Nik09,Li09,Del10,Li10,Li11}
and fission dynamics \cite{WERP02,Bar11,SMBDNS13}.

\begin{figure}
\begin{centering}
\includegraphics[width=0.90\columnwidth]{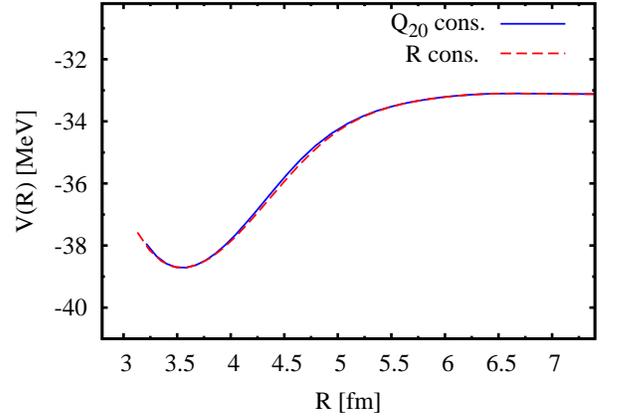}
\par\end{centering}
\caption{\label{fig:pot2}(Color online)
The collective potential obtained with the CHF calculation.
The solid (blue) and dashed (red) lines indicate the results
with constraints on $\hat{Q}_{20}$ and $\hat{R}$, respectively.
}
\end{figure}
\begin{figure}
\begin{centering}
\includegraphics[width=0.90\columnwidth]{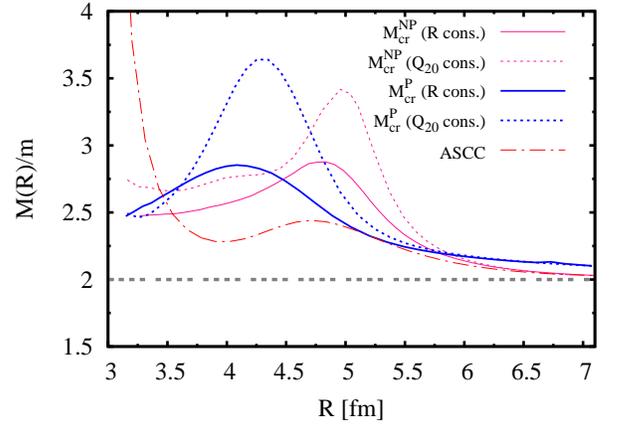}
\par\end{centering}
\caption{\label{fig:mass_cranking}(Color online)
Cranking inertial mass based on the CHF state.
The solid and dashed lines indicate the results
with constraints on $\hat{R}$ and $\hat{Q}_{20}$, respectively.
The non-perturbative and perturbative cranking inertial masses
are shown with thin and thick lines, respectively.
}
\end{figure}

The obtained potentials with different constraint operators
are shown in Fig. \ref{fig:pot2}.
The two constraints $\hat{Q}_{20}$
and $\hat{R}$ give very similar potential surfaces,
which is also close to the ASCC result.
On the other hand, the inertial masses are more sensitive to
the difference.
In Fig.~\ref{fig:mass_cranking},
we show the perturbative and non-perturbative cranking inertial masses
based on the states obtained with CHF calculations with
different constraint operators.
We include all the single-particle states in the model space for the
calculation of Eqs. (\ref{NP_cranking}) and (\ref{S_k}).
They present significant variations, especially in the region where
two $\alpha$'s stick together into one nucleus.
First of all, they are larger than the ASCC inertia.
The second, the non-perturbative and perturbative cranking inertial masses
are significantly different.
For instance, the calculations with $\hat{Q}_{20}$ constraint suggest
prominent peak structure in $M_{\rm cr}^{\rm NP(P)}(R)$.
However, the peak positions are very different.
It should be noted that the present results should not be generalized
to other energy density functionals, because the BKN energy density functional has
no time-odd mean fields (see Eq. (\ref{BKN0})).

Since there are neither effective mass nor time-odd mean field
in the BKN energy density functional, we expect that in the asymptotic region
the exact translational mass $Am$ can be reproduced. 
This turns out to be true for
$M_{\rm cr}^{\rm NP}(R)$, which reduces to the exact value $2m$, while
$M_{\rm cr}^{\rm P}(R)$ approaches to $2m$ much slower than $M_{\rm cr}^{\rm NP}(R)$
and might converge to a larger value.
In fact, for a single $\alpha$ particle,
the translational mass is calculated as $M_{\rm cr}^{\rm P}=4.16m$.
The same kind of deviation is presented in the asymptotic value
of the reduced mass in Fig. \ref{fig:mass_cranking}.


\subsubsection{CHF + local RPA}

Since the cranking inertial mass has known weak points, namely,
missing residual correlations and adiabatic assumption.
The problem becomes particularly serious when the time-odd
mean fields play a role as residual fields.
Although the BKN energy density functional adopted in this paper
does not have the time-odd components,
it may be useful to investigate the significance of the residual effect.

In order to take into account the residual effect,
we adopt the method called ``CHF + local RPA''.
This is defined by replacing $\hat{H}_{\rm mv}(q)$ in
the ASCC equations (\ref{chf}), (\ref{ASCC1}), and (\ref{ASCC2}),
with the constrained Hamiltonian,
$\hat{H}'\equiv \hat{H} - \lambda \hat{O}$,
where $\hat{O}$ is an adopted constraint operator.
In other words,
the collective path is defined by hand, but the inertial mass is
defined by the RPA equations with $\hat{H}'$.
The calculated inertial mass $M_{\rm lrpa}(q)$ for the motion
along the coordinate $q$, can be mapped onto the variable $R$,
$M_{\rm lrpa}(R)$,
assuming the one-to-one correspondence exists between $q$ and $R$.
This is done exactly in the same way as the ASCC (Sec.~\ref{sec:mapping}).
However, the consistency between the generators, $\hat{Q}(q)$ and $\hat{P}(q)$,
and the collective path $\{ \ket{\psi(q)} \}$ is lost.
This method of CHF + local RPA has been applied to studies of
nuclear structure with the separable Hamiltonian
\cite{HSNMM10,HSYNMM11,YH11,SH11,HLNNV12,SHYNMM12}.

\begin{figure}
\begin{centering}
\includegraphics[width=0.90\columnwidth]{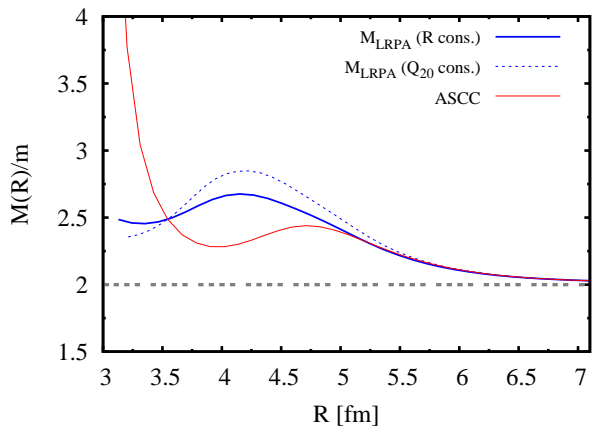}
\par\end{centering}
\caption{\label{fig:mass_LRPA}(Color online)
Inertial mass calculated with the CHF + local RPA in units of nucleon mass.
The solid and dashed lines indicate the results
with constraints on $\hat{R}$ and $\hat{Q}_{20}$, respectively.
The ASCC result is shown by the thin line for comparison.
}
\end{figure}

In Fig.~\ref{fig:mass_LRPA}, we show the result of the local RPA calculation
based on the CHF states.
At the ground state $(R = 3.54$ fm),
since both the CHF + local RPA and the ASCC calculations reduce
to the HF + RPA calculation,
they produces the identical inertial mass.
$M_{\rm lrpa}(R)$ also converges to the ASCC value at large $R$, faster than
$M_{\rm cr}^{\rm NP}(R)$, and asymptotically gives the exact reduced mass $2m$.
Especially, the calculation with the $R$ constraint produces
almost identical results as the ASCC method, at $R>5$ fm.

\begin{figure}
\begin{centering}
\includegraphics[width=0.90\columnwidth]{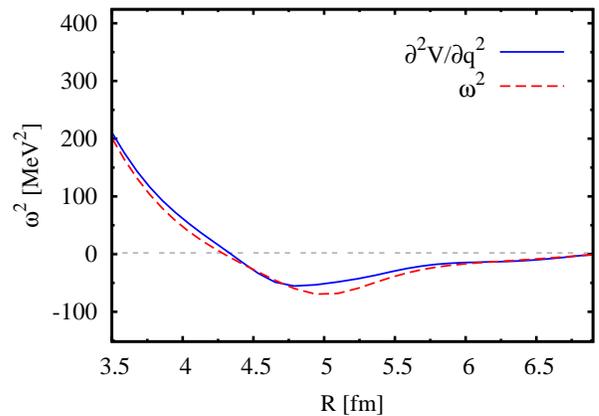}
\par\end{centering}
\caption{\label{fig:checkcy}(Color online)
$\omega^{2}$ in Eq. (\ref{ab3}) and
$\partial^{2}V/\partial q^{2}$ of the CHF + local RPA calculation.
}
\end{figure}

The self-consistency between the local generators and the
assumed coordinate can be checked by comparing the local RPA frequency
and the second derivative of the potential $V(R)$.
If they are consistent, we expect the relation
\begin{equation}
\omega^2 = \frac{d^2V}{dq^2}
=\frac{d^{2}V}{dR^{2}}\frac{1}{M_{\rm lrpa}(R)}
 + \frac{dV}{dR}\frac{d^2R}{dq^2} .
\end{equation}
It turns out that the last term is negligible.
Taking the potential $V(R)$ of the $Q_{20}$ constrained calculation
as an example,
this comparison is shown in Fig.~\ref{fig:checkcy}.
We can see some deviations in the region of $3.5$ fm$ < R < 6$ fm,
although the overall agreement is not so bad.
The deviation indicates that the CHF states are
not exactly on the collective path defined by the local generators
($\hat{Q}(R),\hat{P}(R)$).
On the other hand, the perfect agreement is seen in a region of $R>6$ fm.
This suggests that, at $R>6$ fm, the optimal collective coordinate $q$
obtained with the ASCC method coincides with the relative distance $R$
and the quadrupole moment $Q_{20}$.

Finally, we remark a necessity to modify the constraint operators,
such as $\hat{Q}_{20}$ and $\hat{R}$, in the CHF calculation.
Taking the constraint operator $\hat{Q}_{20}$ as an example, on
the symmetry axis ($z$ axis), the constraint term
$-\lambda\hat{Q}_{20}$ results in a external potential proportional to $-z^{2}$.
If we adopt a large model space, the CHF calculation may lead to
an unphysical solution, namely, the appearance of small density at the
edge of the box.
In order to avoid these unphysical states, we have to screen
the operator in the outer region;
$\tilde{Q}_{20}\equiv
\int f(r) r^2 Y_{20}(\hat{r})\hat{\psi}^\dagger(\vec{r})\hat{\psi}(\vec{r})
 d\vec{r}$
with a screening function $f(r)$ which
should be unity in the relevant region and vanish in the irrelevant region
($r>R_0$).
The function form of $f(r)$ becomes non-trivial when two nuclei are far away
in an asymptotic region.
This kind of complication is not necessary for the ASCC local generator
$\hat{Q}(q)$, because it vanishes in a region
where all the hole orbits are zero $\varphi_i(\vec{r};q)=0$.
In other words, the ASCC generators are properly ``screened'' automatically.

\subsubsection{ATDHF}

The ATDHF is based on Eqs. (\ref{chf}) and (\ref{ASCC1}).
Since the second-order equation (\ref{ASCC2}) is missing,
the collective path is not unique.
We follow the prescriptions given in Ref.~\cite{RMG80}
for practical calculations.
The equation of the collective path
is formulated in a form of the first-order differential equation for
$|\psi(q)\rangle$,
\begin{eqnarray}
\frac{\partial}{\partial q}|\psi(q)\rangle
 = \frac{M_{\rm atdhf}(q)}{d V/d q}[\hat{H},\hat{H}_{\rm ph}]_{\rm ph}|\psi(q)\rangle,
\label{atdhf0}
\end{eqnarray}
where $\hat{H}_{\rm ph}$ is the ph and hp parts of the Hamiltonian
defined locally at each $q$.
The single-particle wave functions $\ket{\varphi_i(q)}$ in the
Slater determinant $\ket{\psi(q)}$ is evolved according to
the following equation:
\begin{eqnarray}
\ket{\varphi_i(q-\delta q)} &=& \ket{\varphi_i(q)}
 - \varepsilon \left\{ 1-\rho(q)\right\} \nonumber\\
& \times& \left(
h_{\rm HF}(q) \left\{1-2\rho(q)\right\} h_{\rm HF}(q)
 \right. \nonumber \\
&&\left.
+ \textrm{Tr} \left\{v [h_{\rm HF}(q),\rho(q)] \right\}
\right)
 \ket{\varphi_i(q)}
\label{atdhf}
\end{eqnarray}
with
\begin{eqnarray}
\varepsilon &=& \frac{\delta q M_{\rm atdhf}(q)}{d V/d q}.
\label{atdhfm2}
\end{eqnarray}
In order to
obtain the stable solutions, $\varepsilon$ is set to be a small real number.
Successive application of Eq. (\ref{atdhf}) gives the ATDHF collective
path.
The solutions with different initial
conditions of $|\psi(0)\rangle$ produce
different collective paths.
The envelope curve of all these
trajectories is regarded as the final solution of the adiabatic collective path.

The ATDHF inertial mass is given by
\begin{eqnarray}
M_{\rm atdhf}(q) = \langle \psi(q) | [\hat{Q}(q),[\hat{H},\hat{Q}(q)]] | \psi(q) \rangle^{-1},
\end{eqnarray}
with
\begin{eqnarray}
\hat{Q}(q) = \left(\frac{\partial V}{\partial q}\right)^{-1} \hat{H}_{\rm ph}(q)
= \left(\frac{\partial V}{\partial q}\right)^{-1}
 \left\{ h_{\rm HF}(q) \right\}_{\rm ph}.\nonumber\\
\end{eqnarray}
According to Eq. (\ref{mass}), the mass with respect to the relative
distance $R$ can be calculated as
\begin{eqnarray}
&&M_{\rm atdhf}(R) = M_{\rm atdhf}(q)\left(\frac{dq}{dR}\right)^{2} \nonumber\\
           &=&\left(\frac{d V}{d R}\right)^{2}
           \langle \psi(q) | [\hat{H}_{\rm ph}(q),[\hat{H},\hat{H}_{\rm ph}(q)]] | \psi(q) \rangle^{-1}.\nonumber\\
           \label{atdhfm1}
\end{eqnarray}
Another, even easier, way of calculating $M_{\rm atdhf}(R)$ is simply
inverting Eq. (\ref{atdhfm2}).
Using Eqs. (\ref{mass}) and (\ref{atdhfm2}),
we obtain
\begin{eqnarray}
M_{\rm atdhf}(R)= \left(\frac{dq}{dR}\right)^{2} \frac{\varepsilon}{\delta q}\frac{d V}{d q}
      = \frac{\varepsilon}{\delta R}\frac{d V}{d R}.
\label{atdhfm3}
\end{eqnarray}

%

\begin{figure}
\begin{centering}
\includegraphics[width=0.90\columnwidth]{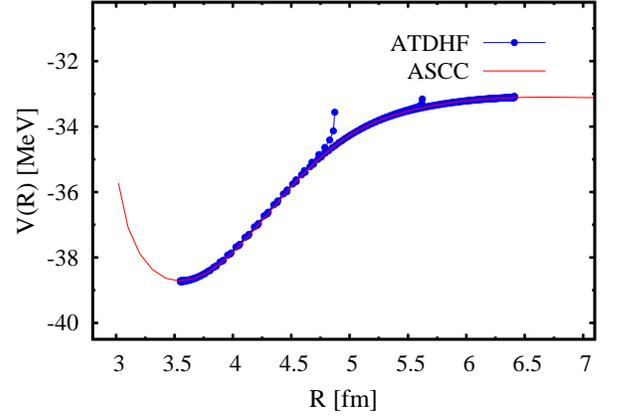}
\par\end{centering}
\caption{\label{fig:atdhfv}(Color online)
The potential energy on the ATDHF collective path derived
by Eq. (\ref{atdhf}), as a function of relative distance $R$.
Initial distances between the two alpha particles are set to be
$R = 4.8, 5.6, 6.4$ fm respectively.
The thin (red) line indicates the result
of ASCC method.
}
\end{figure}

For the scattering between two $\alpha$'s,
we prepare two $\alpha$ particles both
at ground states separately,
then put them away at different distances of $R = 4.8$, 5.6, 6.4 fm,
as the initial conditions for Eq. (\ref{atdhf}).
The potential surface of the ATDHF trajectories are
plotted in Fig.~\ref{fig:atdhfv},
which shows how the solutions of Eq. (\ref{atdhf0}) with different
initial conditions converge to
a common collective path. 
The converged ATDHF potential
surface is similar to the potentials of CHF and ASCC calculations.
It should be noted that we can obtain these
fall-line trajectories on the potential surface
which go only from high to low energy \cite{RMG80}.
It becomes numerically unstable if we calculate in the opposite direction.
Thus, we cannot start from the HF ground state, and
it is difficult to obtain the solution in a region of $R<3.5$ fm,
beyond the HF minimum state.
\begin{figure}
\begin{centering}
\includegraphics[width=0.90\columnwidth]{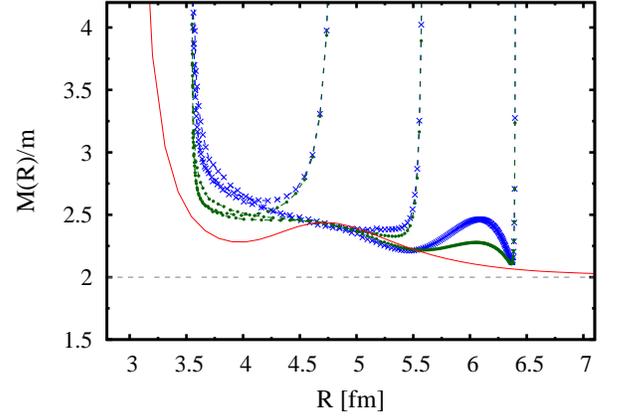}
\par\end{centering}
\caption{\label{fig:atdhfm}(Color online)
$M_{\rm atdhf}(R)$ calculated by Eqs. (\ref{atdhfm1})
and (\ref{atdhfm3}) shown with blue crosses and green dots, respectively.
They are based on the
same ATDHF trajectories in Fig. \ref{fig:atdhfv}.
The solid (red) line indicates the ASCC mass for comparison.
}
\end{figure}

Figure \ref{fig:atdhfm} shows the mass parameters
based on the
same trajectories in Fig. \ref{fig:atdhfv}.
The inertial masses calculated with Eqs. (\ref{atdhfm1}) and (\ref{atdhfm3})
roughly produce the identical results.
Near the HF state of $R = 3.54$ fm,
the inertial mass increases drastically.
This is very different from the result of the former calculations
\cite{RMG80,GGR83},
the reason of which is currently under investigation.
We also encounter a difficulty to obtain the collective path in the
asymptotic region at large $R$.
A larger model space and finer mesh size seems to be needed
to obtain the potential in the asymptotic region
and to reproduce the reduced mass $2m$.
We should also mention that the saddle point with $dV/dR=0$ is
extremely difficult to reach by solving Eq. (\ref{atdhf}).
In the ASCC method, we do not encounter these difficulties,
and are able to obtain the unique reaction path and inertial mass.

\section{Summary}
\label{sec:summary}

We have applied the ASCC method to the determination of the nuclear reaction
path, the collective potential, and the collective inertial mass.
The 3D coordinate space representation is adopted for the single-particle
wave functions.
Using the imaginary-time method
and the finite-amplitude method, the coupled equations of the ASCC,
that consist of the moving HF equation and the moving RPA equations,
are solved iteratively.
The generators are represented in the mixed representation of
the hole orbit and the coordinate grid points, such as $Q_j(\vec{r})$.

The first application has been performed to the simplest case,
the scattering of $\alpha+\alpha \leftrightarrow ^8$Be.
The reaction path, the potential, and the inertial mass are successfully
determined.
Even though the system is too simple to expect significant difference in
the reaction path,
a comparison with the cranking inertial mass demonstrates some advantages
of the ASCC method.
In particular, the cranking inertial mass is very sensitive to
the adopted prescription of perturbative or non-perturbative formulae.
The perturbative cranking mass seems not reduce to the
exact value of the reduced mass at $R\rightarrow\infty$.
For $^8$Be, the potential does not depend on the choice of the
constraint operator.
In contrast, a proper choice of the operator is important for the
inertial mass.
The ASCC method is able to remove these ambiguities and provide
improvement of the cranking formula.
The ATDHF theory is an alternative way to derive the reaction path and
inertial mass.
However, we have found that to find the unique converged result of
the ATDHF trajectories is significantly more difficult than the ASCC method.

The reaction path and the feature of the inertial mass
depend on the reaction system.
The calculation for heavier systems is under progress.
With the techniques presented in this work,
it is feasible to perform the calculation of the inertial mass
for different modes of nuclear collective motion,
such as the rotational moment of inertia, and
the mass parameter for different vibrational modes.
The lowest mode of excitation changes from nucleus to nucleus,
and we shall investigate how these nuclear excitation properties
influence the reaction dynamics.

The simple BKN energy density functional should be replaced by
a modern nuclear energy density functional, in future.
The presence of time-odd mean fields would significantly affect
dynamical behaviors of nuclear systems.
Since the cranking inertia cannot take into account the time-odd effects,
advantages of the ASCC method become even clearer.
The inclusion of the paring correlation is another important issue.
This has been studied in nuclear structure problems \cite{NMMY16}.
However, for the nuclear reaction studies, some conceptual problems
for the paired systems still remain to be solved.
For instance, the ASCC method for
the reaction of two nuclei with different chemical potentials
has not been established yet.
This is also an important subject in future.

\begin{acknowledgments}
This work is supported in part by JSPS KAKENHI Grants No. 25287065
and by ImPACT Program of Council for Science,
Technology and Innovation (Cabinet Office, Government of Japan).
\end{acknowledgments}

\bibliography{myself,nuclear_physics,current,current2}


\end{document}